# The Oracle of Chemnitz: An interactive art installation to reanimate old things in a garage featuring a rotary phone

**Karola Köpferl**
Chemnitz University of Technology
Chair for Ergonomics and Innovation Management
Chemnitz, Germany
karola.koepferl@mb.tu-chemnitz.de

**Albrecht Kurze**
Chemnitz University of Technology
Chair Media Informatics
Chemnitz, Germany
albrecht.kurze@informatik.tu-chemnitz.de

## Abstract
Garages have a long tradition of tinkering, creativity and innovative change. School of Garage, a participatory artistic summer school project in Chemnitz, the European Capital of Culture 2025, took up this tradition and turned old Eastern Bloc garages into temporary ateliers for collaborative making and discussion. In our HackLab garage we conceptualized and created the Oracle of Chemnitz within one week. It gives a place filled with history back its stories. It is an interactive installation of artifacts from the past typically found in garages: an old typewriter, radio, desk, tires, mixer and a rotary-dial telephone. Each got a name, personality and story to tell. The phone rings when a visitor approaches. Once answered, it asks for name and month of birth before a story about a device is told, along with hints to other places in the city. Around 2,700 visitors interacted with the system over three months.

## Introduction
Garages have a long tradition of tinkering, creativity, and innovative change. In Western culture, one might think of them as spaces for start-ups and creative activities, such as music and art (Erlanger & Govela, 2018). Nowadays, even large companies consider garages to be innovation hubs, places where people can work together intensively in in proximity (Kamiński & Rosłoń, 2024).

In the Eastern Bloc, despite a lower rate of cars per capita, garages have always played an important role in addition to providing car parking. Despite their standardized architecture and organization in large garage yards, Eastern garages also have a long tradition of being places for repair and storing parts, as well as for creativity and socializing with neighbors (Crowley & Reid, 2002; Tuvikene, 2011).

*What if garages from the era when Europe was divided by the Iron Curtain are reanimated and became sites of experimentation, encounters, and new meanings?*

We transformed a garage into a space for dialogue, where discarded items are given new roles and grassroots collaboration is fostered through creative processes. Developed within the concept of Chemnitz European Capital of Culture 2025, which positions the city as a hub for a shared culture of making with the aim of fostering self-efficacy and civic engagement, the project was part of the School of Garage summer school. Here, garages were explored as contemporary laboratories for collaborative artistic and technological experimentation. The resulting installation reactivates familiar garage artefacts through digital technologies, translating everyday practices into a public experience. Our work resembles techno-artistic and aesthetic approaches and practices (Bødker, 2006), including critical, reflective, and speculative design and making (Dunne & Raby, 2013), materially-engaged and multisensory interactions (Frankjær & Dalsgaard, 2018), and multidisciplinary collaborations (Kang et al., 2018).

Our interactive installation builds on salvaged old artefacts (rotary phone, mechanical typewriter, mixer, radio, tires, drawer-desk) that can be found stored and often forgotten in garages. Visitors interact with the system through voice dialogs with the telephone acting as an oracle that gives the artifact voices and let them tell stories of their past. The installation thus combines retro technology, tech nostalgia and storytelling about events, places and relationships – aspects that are associated with local history.

## Chemnitz – European Capital of Culture 2025
Chemnitz, located around 200 km south of Berlin, is the 4th largest city in eastern Germany with around 245.000 inhabitants. The city has a rich cultural and industrial heritage. In the middle age it played an important role as gate to Ore Mountains in the south, was considered as "Saxon Manchester" in the industrialization in the 19th century and is called "City of the Modern" in art. In World War II the city was widely destroyed. During the Cold War it was behind the Iron Curtain in the "German Democratic Republic" (GDR) and named "Karl-Marx-Stadt". In this time large parts of the city were rebuilt in eastern communist style with lots of concrete and uniform architecture. After the wall came down in 1989 the city was renamed back to Chemnitz in 1990.

In 2025, Chemnitz served as European Capital of Culture under the motto #Chemnitz2025: "C the Unseen", highlighting overlooked places, practices, and everyday cultures that have shaped the city despite often remaining invisible at a national and international level.

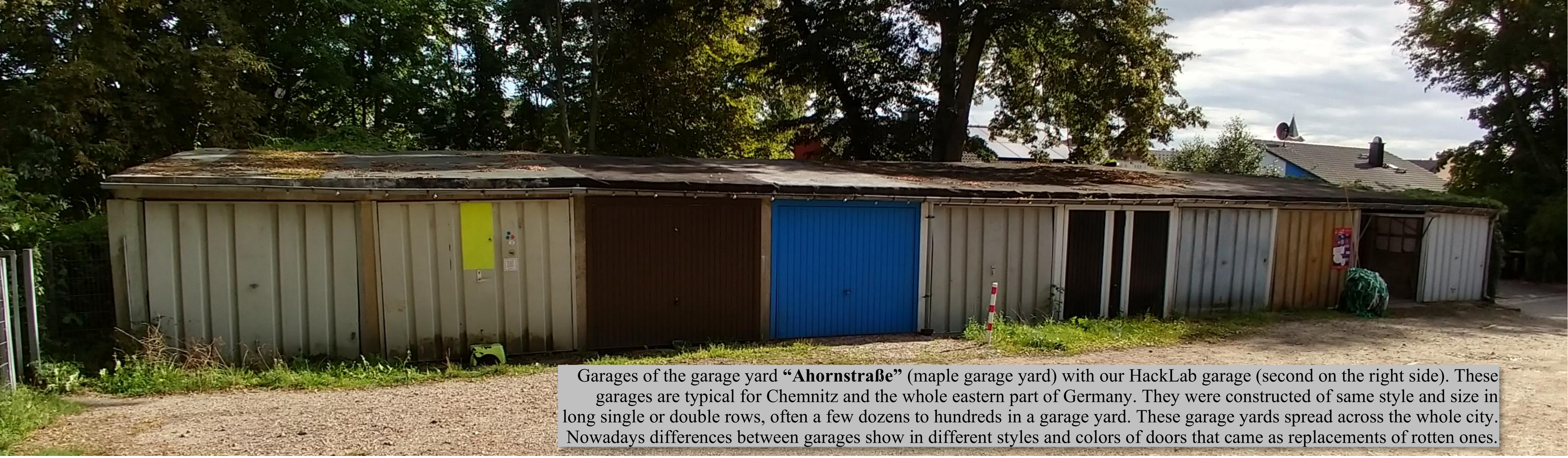

Garages of the garage yard **“Ahornstraße”** (maple garage yard) with our HackLab garage (second on the right side). These garages are typical for Chemnitz and the whole eastern part of Germany. They were constructed of same style and size in long single or double rows, often a few dozens to hundreds in a garage yard. These garage yards spread across the whole city. Nowadays differences between garages show in different styles and colors of doors that came as replacements of rotten ones.

## The role of garages

In Chemnitz, where approximately 30,000 garages shape the urban landscape, these spaces played a central role in everyday life during the time in East Germany. Most garages were built collectively by residents and served as places for tinkering, repair, and informal creativity. Beyond parking and storage, garages became sites of repair culture where scarcity fostered hands-on problem solving, improvisation, and technical self-reliance (Krause, 2025; Kulturhauptstadt Europas Chemnitz 2025 gGmbH, 2025; Wehner, 2025). Garages also functioned as social spaces - often described as “extended living rooms” - enabling community interaction, mutual assistance, and private expression under conditions of surveillance. Collective weekend construction and shared use reinforced solidarity while embedding garages deeply in neighborhood life. Today, garages continue to act as interfaces between private and public space. (Bonvin, 2023)

The **#3000Garages** project builds on this legacy by approaching garages as cultural heritage and living archives. Through artistic, social, and participatory formats, the project reactivates garages and garage courtyards as sites of creation, learning, and community interaction. In doing so, it highlights a shift from Eastern Bloc GDR-era utility toward contemporary socio-cultural potential, positioning garages as spaces where individual histories and collective practices intersect (Kulturhauptstadt Europas Chemnitz 2025 gGmbH).

## School of Garage and our HackLab

School of Garage was conceived as a participatory artistic summer school in nine garages spread across the city. The garages were planned to become laboratories for ideas, exchange, and collective action by Constructlab (2025). Each garage had its own main initiators and focus, from upcycling, urban gardening, storytelling, solidarity work, painting, space appropriation, favorite mobiles, photo sculpting to technology. An open call for participation invited to apply for one of the garages for a one-week summer school in the end of August 2025. The summer school provided an organizational frame for all participating garages. A kickoff on the starting weekend brought together all participants in a central space for a come together and starting their projects. Then for the next weekdays each garage team worked in their own garage.

Our **HackLab** garage was conceptualized as an open tech studio for coding, building, experimenting and transforming a garage into space for collective interventions. The garage should function as a contemporary laboratory and social hub, where participants learn and share knowledge through doing, repairing, and hacking. Working in an interdisciplinary setting, the group should combine different skills, perspective and biographies while actively involving local actors, including members of the Chaos Computer Club Chemnitz (ChCH).

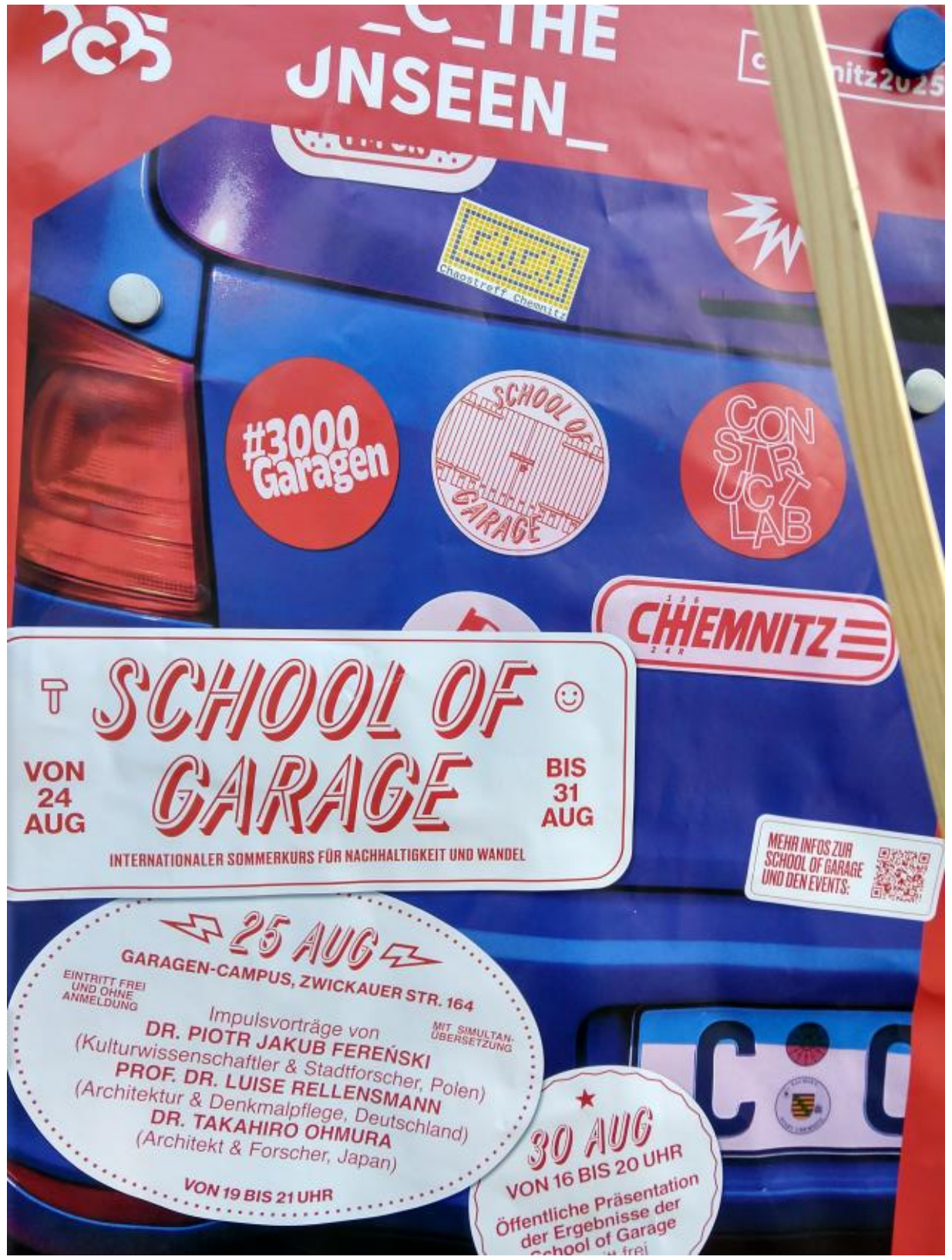

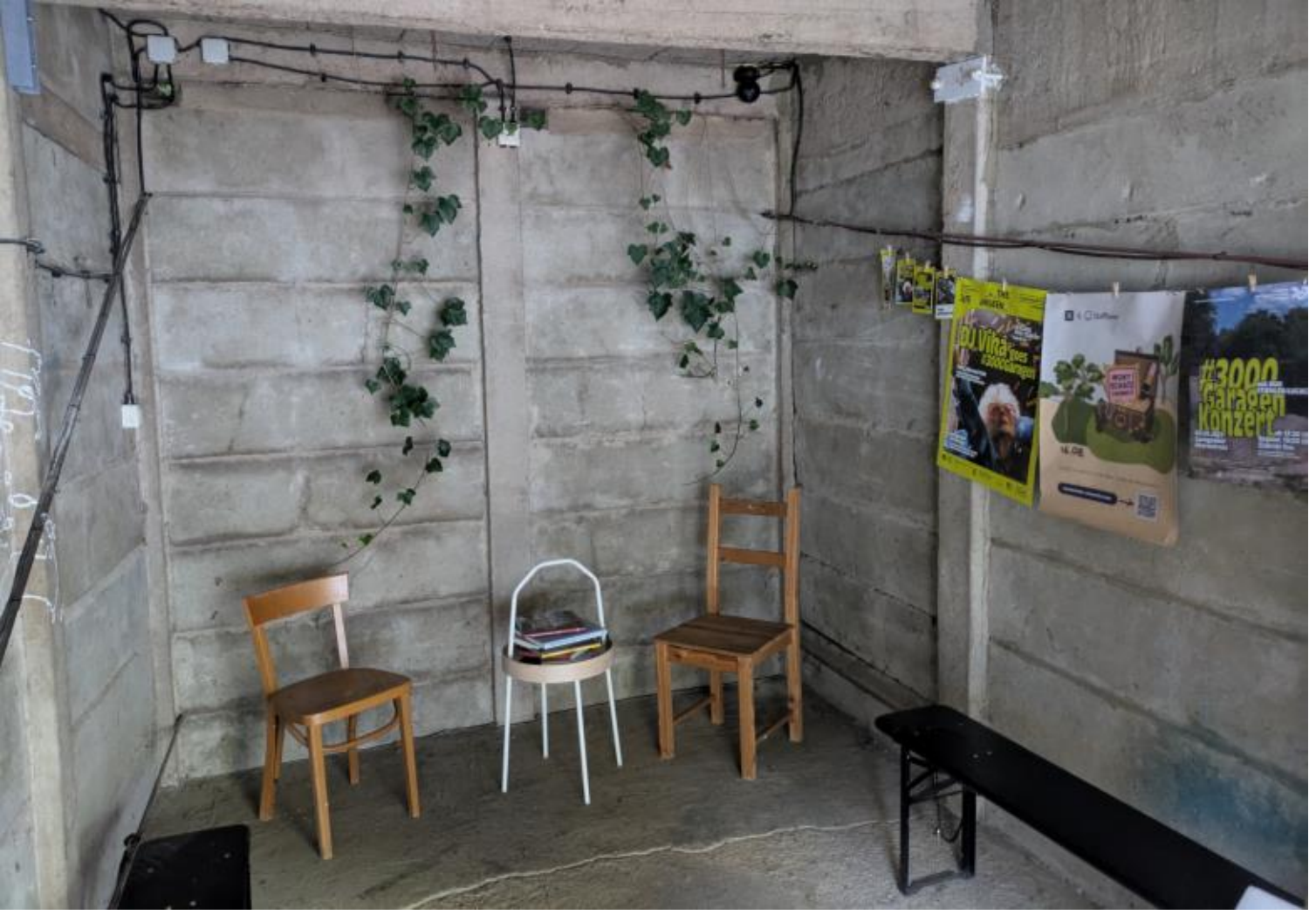

The **HackLab garage team** brought together artists, students, hackers, and local residents with diverse backgrounds in art, architecture, media design, mechanical engineering and computer science. The HackLab core team consisted of eight people (3f, 5m) with age between 25 and 50; with 2 design educators, 4 students, 2 ChCh members. The garage owner (with a background in IT) and some neighbors also became active participants in the process. Other members of the Chaos Computer Club Chemnitz contributed expertise in software, hardware and electronics. Since the team was multinational (CZ, DE, IT, PL) we used mainly English for communication but also German, Italian and Polish. Rather than following a predefined hierarchy, the group worked as colleagues learning from one another, valuing openness, improvisation, and humor as integral parts of the collaborative process. We also worked together with some of the other School of Garage projects.

In the weeks before the summer school, the garage owner had already emptied his garage, which he mainly uses to store his garden furniture in winter, and created a small quiet room to take a rest there. Favored by the warm and sunny late summer weather a lot of the activities were not only performed in the garage itself but also in front of it in the public garage yard. This led to increased visibility of what the team was doing. It attracted some curious tourists but first of all neighbors and sometimes even involved them in ongoing activities.

## Creative process start and material sourcing

The collaborative creative process started with an open brainstorming. The team collected ideas involving plants, about connecting devices, a garage network (analog or digital, to give things new functions or switch functions between things, about participative tools and devices, the whole garage as an instrument, the smells of 70s spirit, about messages in bottles (with NFC). New mythology, amber, storytelling and an oracle that gives advices what to do and where to go came up. Eventually these discussions about values, memories, and everyday objects gradually converged around a rotary-dial telephone and the story of "Aunt Ursula", which became the central narrative anchor of the *Oracle of Chemnitz*. The group envisioned transforming the garage into an oracle-like space, where forgotten machines such as telephones, typewriters, and mixers would be reactivated and given a new, almost magical function (Köpferl & Kurze, 2025a). Each object was imagined as a speaking entity, carrying its own history, memory, and voice.

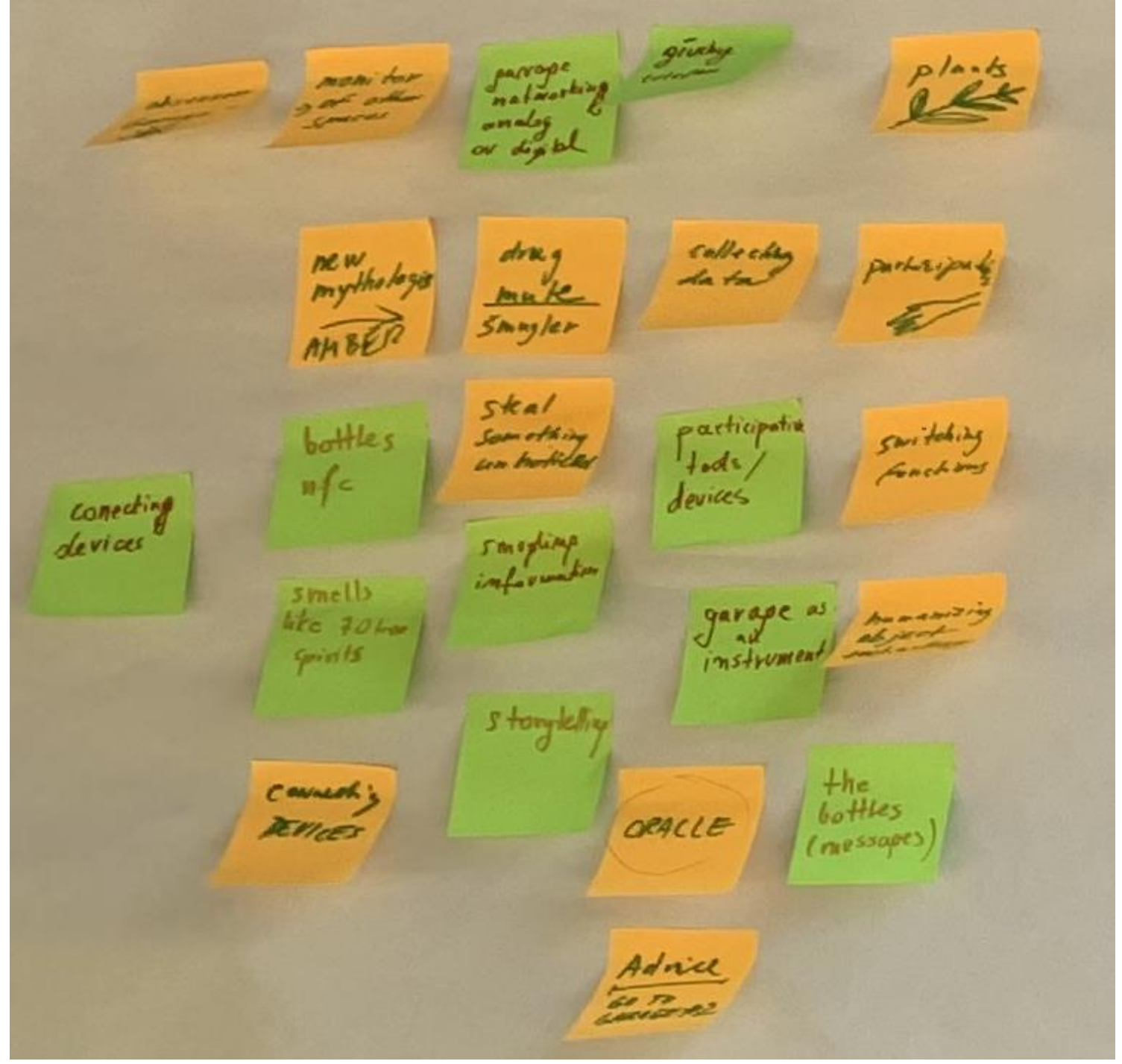

The next steps followed a hands-on approach. Participants sourced **materials and objects** directly from nearby garages and public spaces, reinforcing the site-specific and embedded nature of the work and treating the neighborhood itself as a resource. Eventually this process turned the previously empty garage into a temporary material depot.

Participants sourced old furniture, a heater, old switches and cabling, home and kitchen appliances, a mechanical typewriter, a variety of phones, a tv, a radio, some books, some warning and old traffic signs, a wall clock, garden hoses, oilcans, and several decorative items etc.

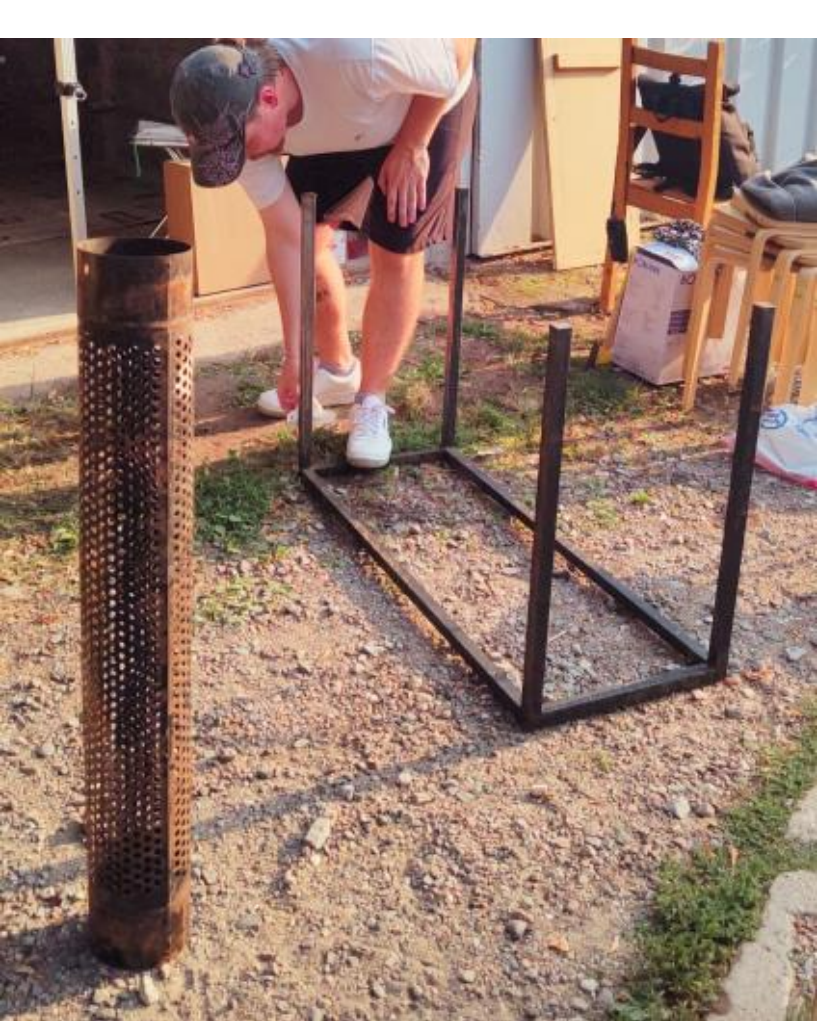

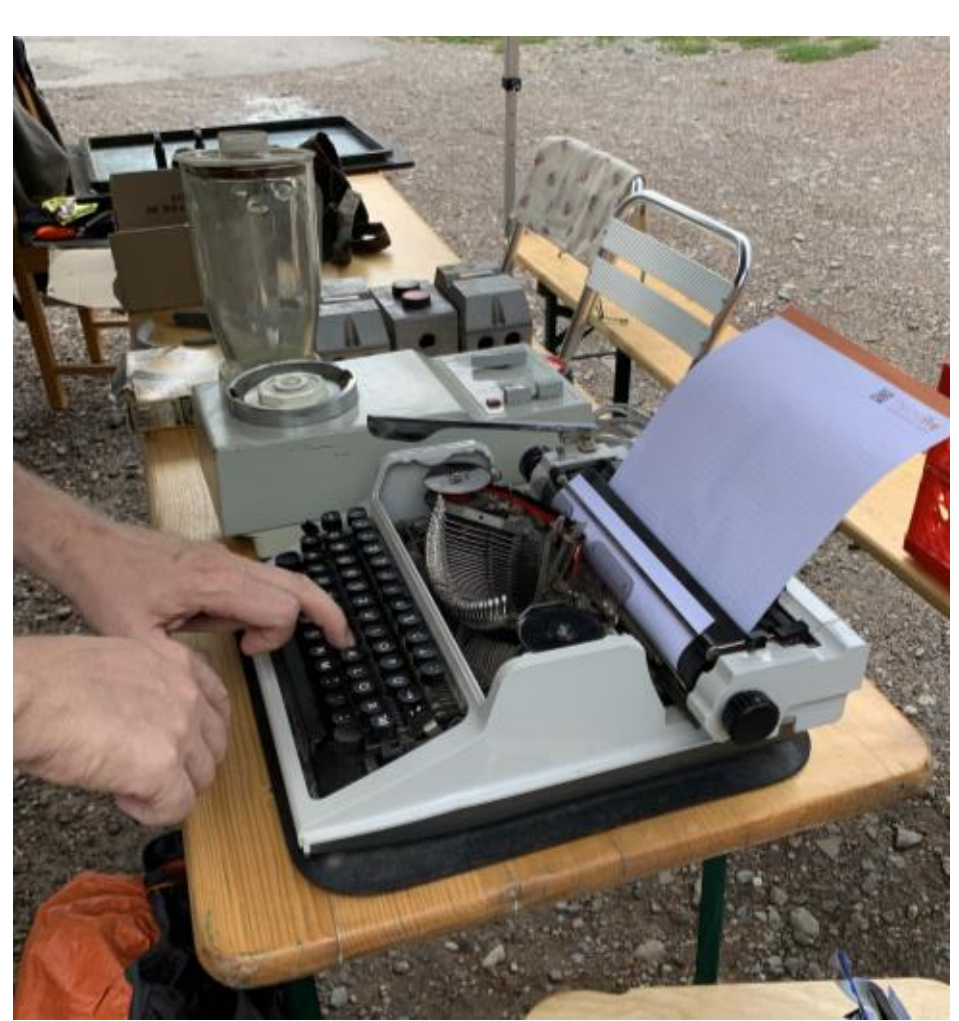

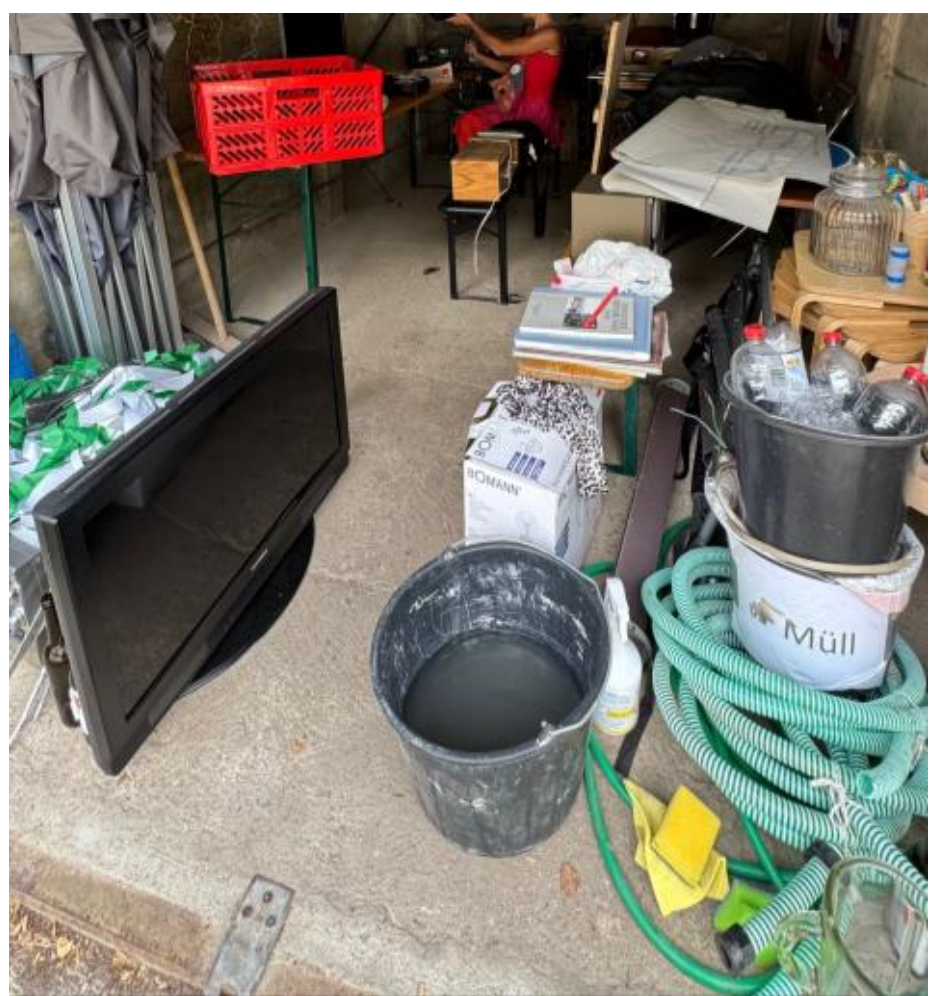

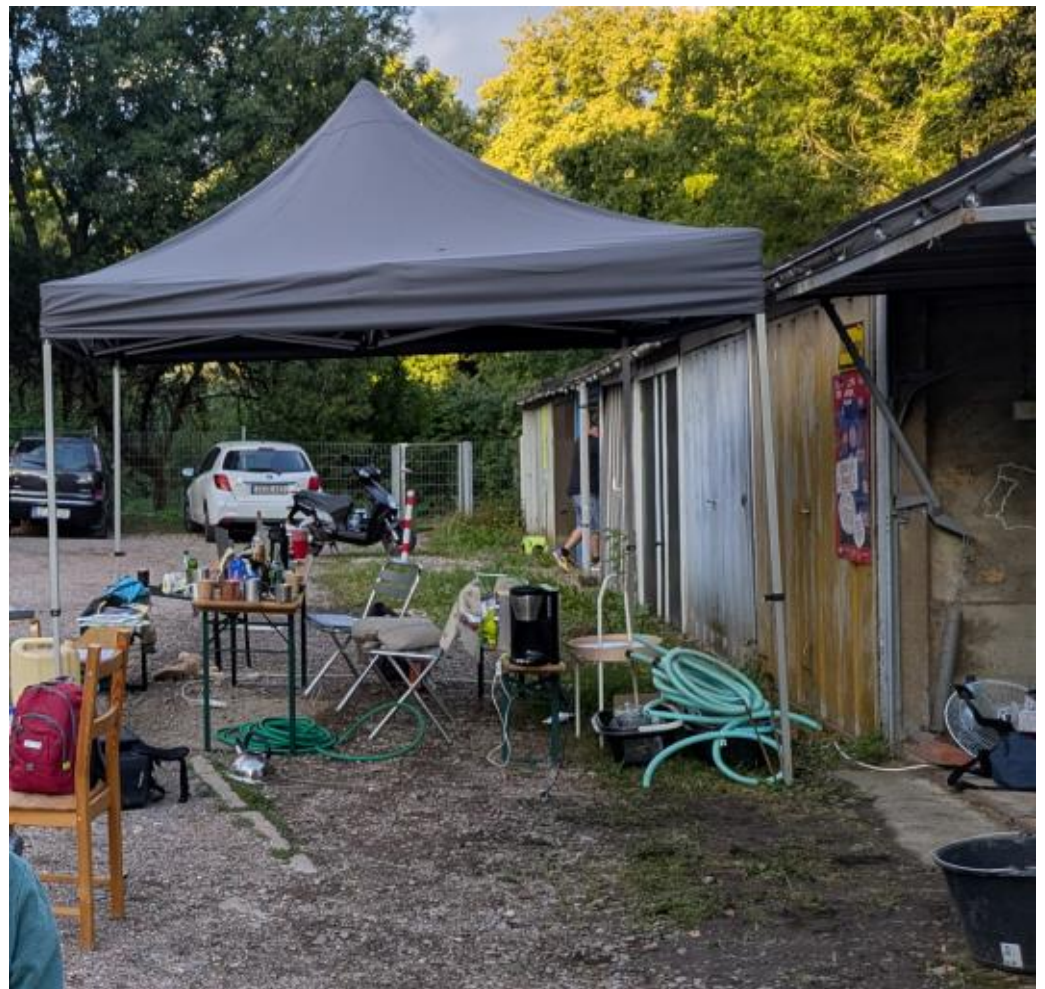

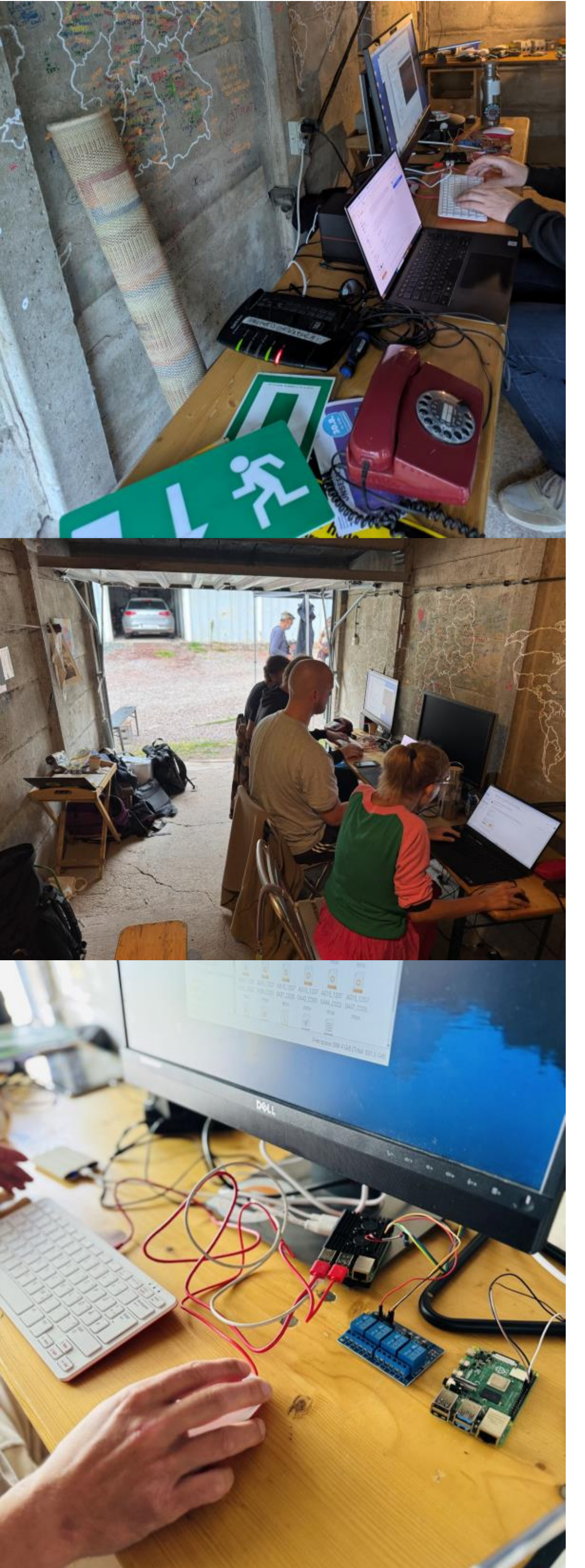

## Making, building and hacking our oracle

In the following iterative process, we combined experimentation, trial and error, and spontaneous collaboration, allowing technical decisions and narrative elements to evolve together rather than being fixed in advance.

Together with additional maker components donated by team members we got all needed parts: the phone itself as main interface to the users, a Fritz!Box home router as a SIP phone gateway between the analog phone and software driven parts of our system. A Raspberry Pi runs our software, especially bringing the oracle to life and giving it a voice, to control external components via a relay board, and a simple passive infrared (PIR) motion sensor to detect approaching visitors.

As the week progressed, the HackLab garage gradually transformed into a dense working environment. Building, soldering, coding, and testing happened side by side, often simultaneously. Rather than separating conceptual and technical work, ideas were continuously refined through hands-on making. The physical presence of the objects shaped both interaction design and narrative decisions, reinforcing the garage as a space where thinking happens through doing.

Technical challenges, particularly around integrating the rotary-dial telephone, required repeated experimentation and collective problem-solving. A circuit plan for the telephone together with documentation found online, created by a community of retro tech hobbyists, helped us to repair and adapt the phone for our purposes. In this process we disabled the pulse dialing to prevent interference with the system. Once a small collection of potentially interesting artifacts was salvaged, we continued to conceptualized the roles of each artifact and how the all artifacts might be integrated in an interactive installation that fits the theme and frame of the garage.

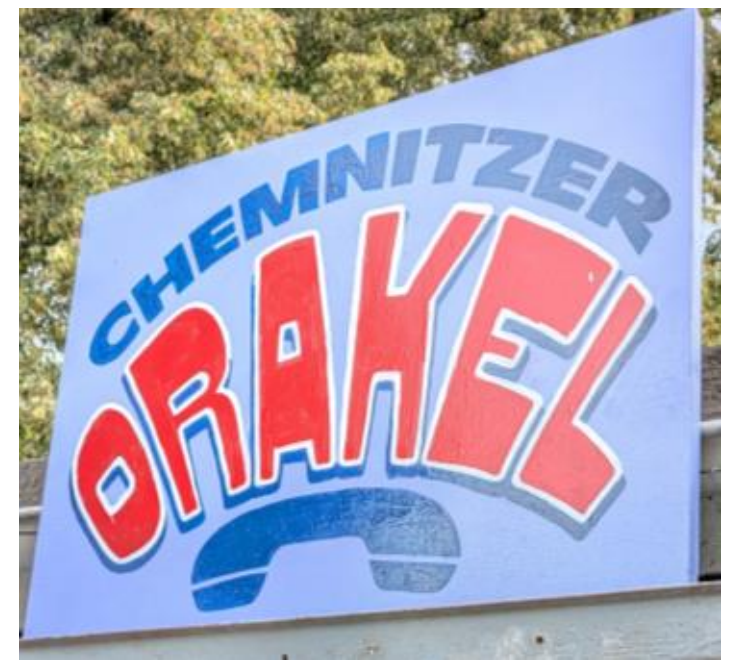

Another garage team made us a big sign that picks up the iconic telephone handset symbol found on old signs.

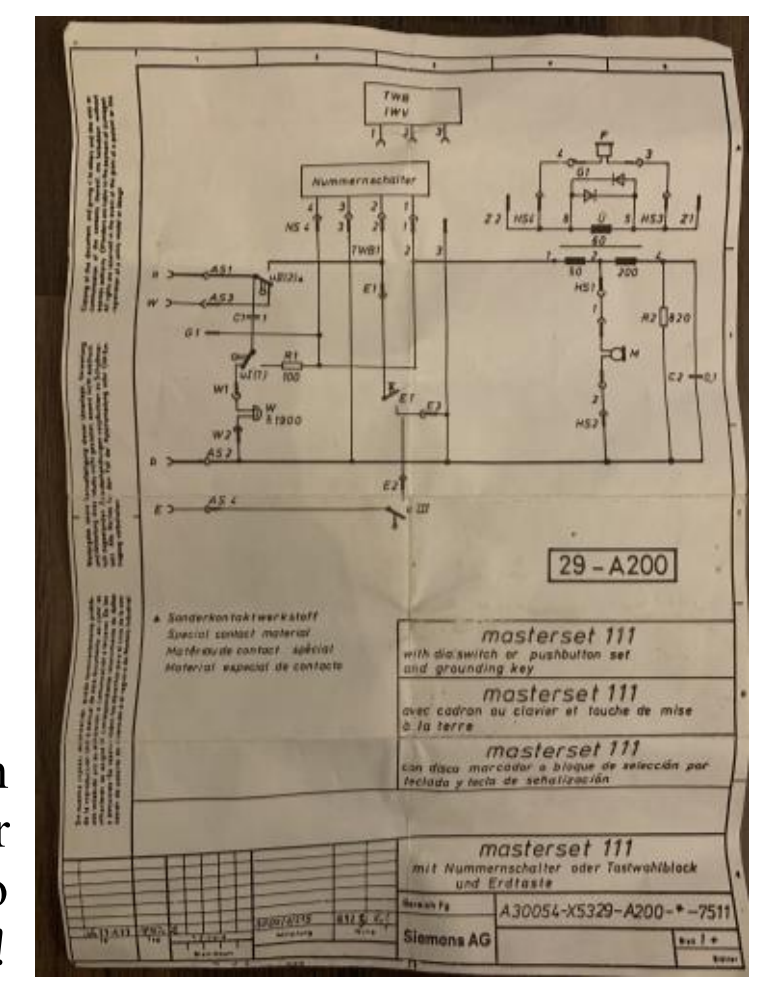

We found a circuit plan directly in folded form in one of the old phones. This was quite a surprise for everybody born in the 80s and later on. Shut out to the engineers and manufacturers of the 70s & 80s!

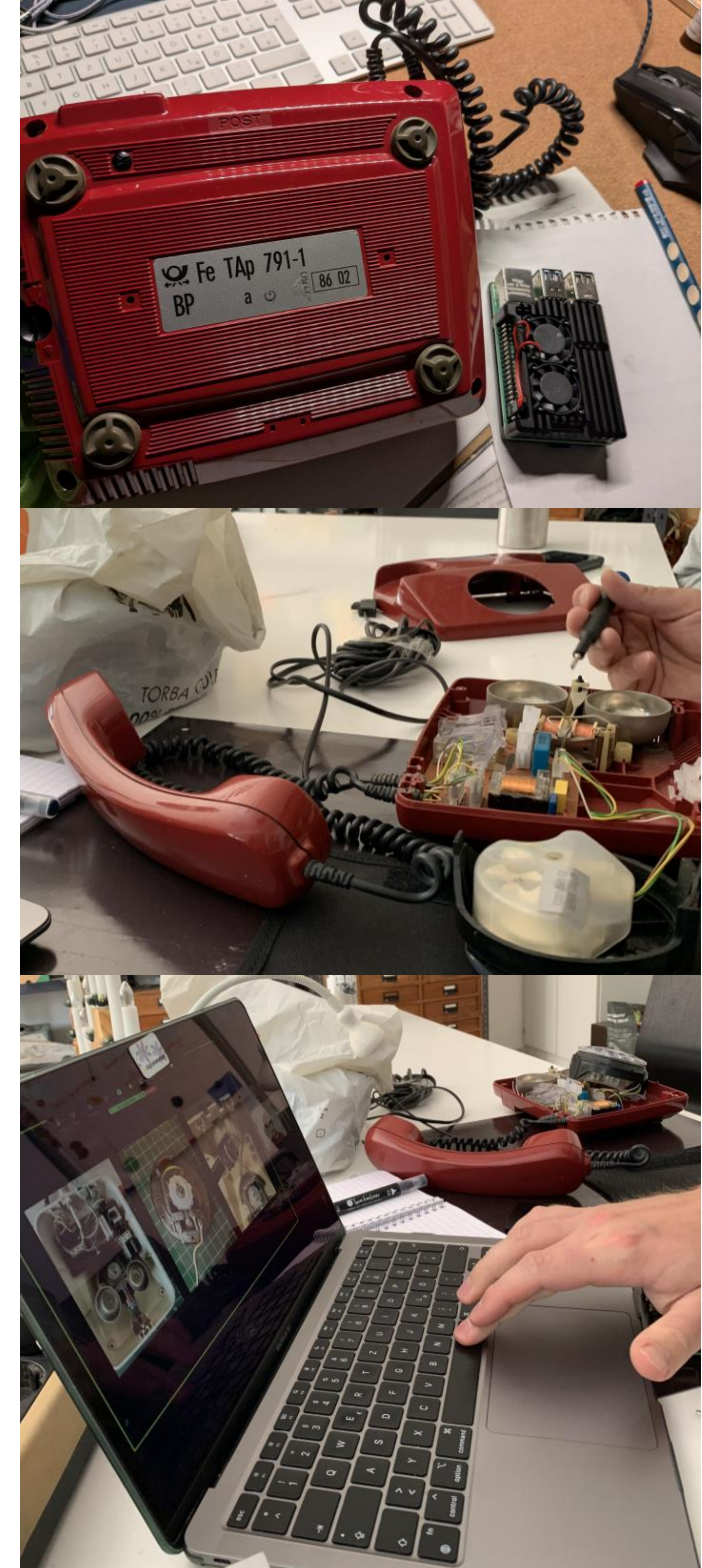

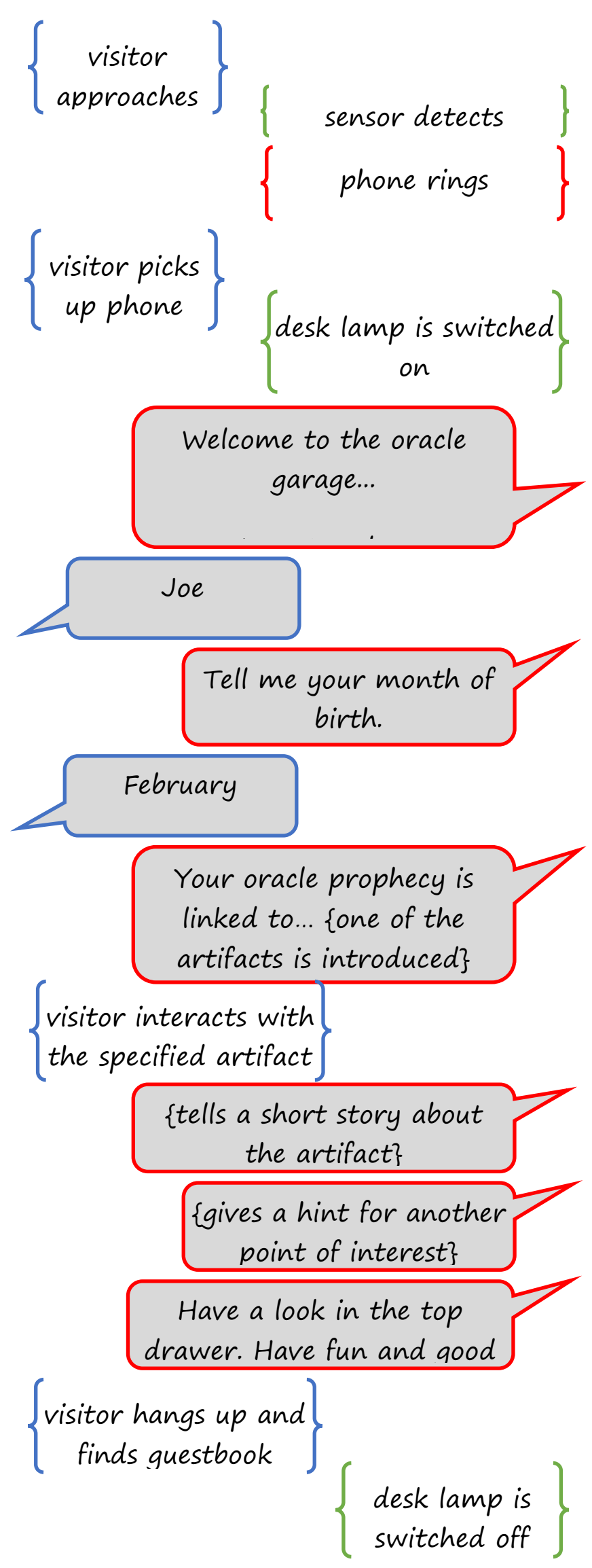

The interaction sequence: spoken text in speech bubbles, other interactions in brackets, visitor interactions in blue on the left, system interactions on the right in green and in red with Ursula, the rotary telephone.

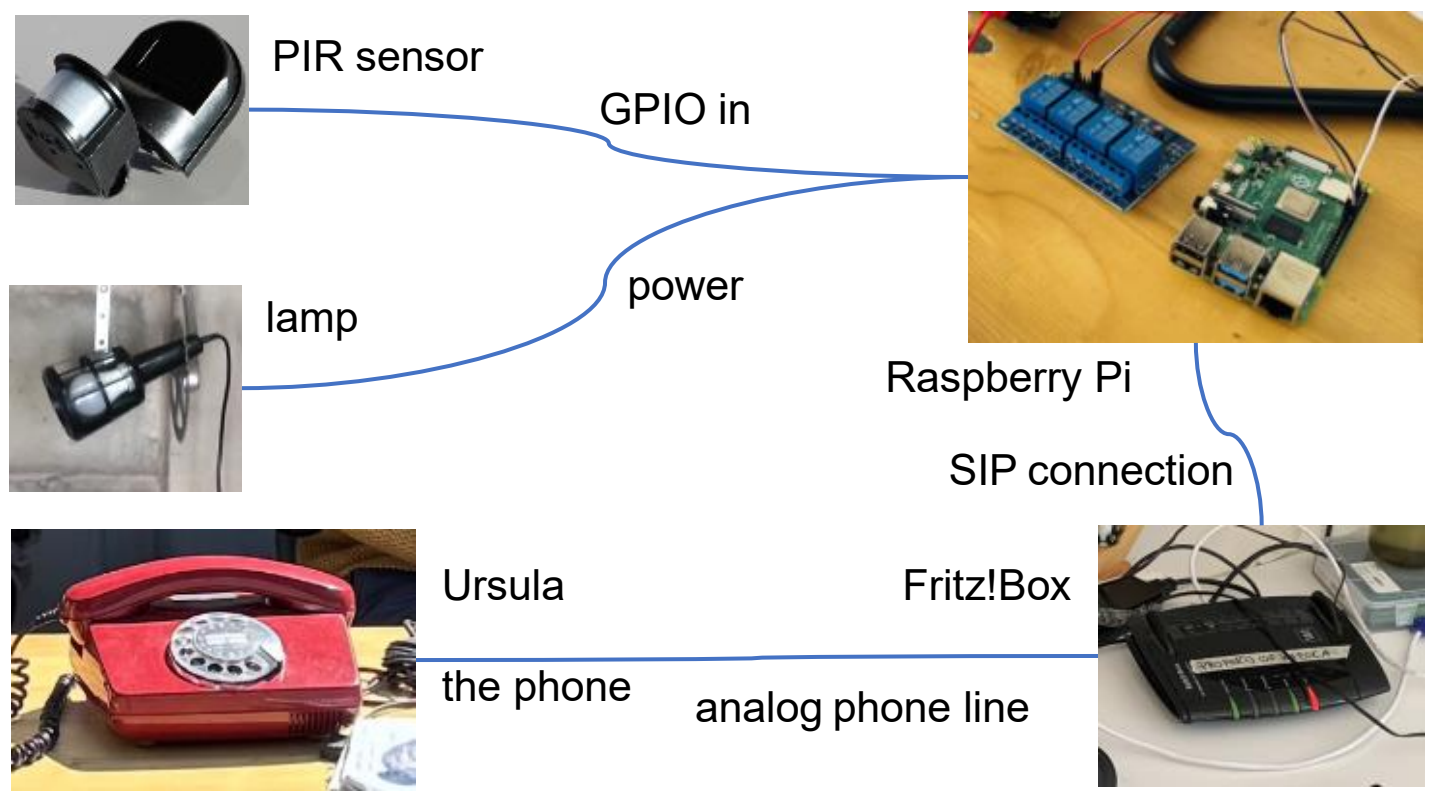

## The interaction concept

We set our focus not on technological novelty, but on creating an intuitive and engaging encounter rooted in familiar objects and gestures. We deliberately chose simple sensing and tangible control mechanisms to keep the interaction legible and accessible for all ages (Bong et al., 2020).

The system is activated by the presence of a visitor, triggering the ringing of the phone. Once picked up, the phone guides the interaction through spoken prompts and connects visitors to different objects within the garage, giving them a voice and a story, initiates further interactions, gives a hint for another point of interest in Chemnitz and finally to the guestbook in one of the drawers of the desk. While the phone call is going on a desk lamp is switched on to focus attention on the installation.

We designed the interaction sequence to take around 1:30 min. This allows visitors to use the oracle in groups without long waiting times or to listen to more than one oracle story. It also leaves enough time to explore the whole installation and all accompanying objects or to read the artifacts background stories.

The interaction sequence is controlled by simple Bash and Python scripts and libraries (e.g., SIP for connecting to the phone) on the Raspberry Pi. The main script implements a simple state machine that processes the inputs and then goes to the next step until the interaction sequence is ended or interrupted by hanging up the phone. We configured the Raspberry to start the whole system directly on bootup without further interaction needed. This makes the system robust, easily usable and in case of a malfunction a simple power-cycle reinitiates everything. This was important since the installation was meant to be operated unsupervised or at best just power-cycled by the garage owner or a neighbor.

## The artifacts and their stories

The installation consists of a set of named everyday artifacts, each associated with a distinct role and character. By assigning names and voices to these artifacts, the installation invites visitors to relate to familiar garage objects as active participants rather than passive exhibits. Ursula, the rotary-dial telephone, functions as the central interface, addressing visitors directly and guiding the interaction. Other objects, such as Erika (a mechanical typewriter) and Manfred (an old mixer), extend the experience through physical engagement and storytelling. Additional artifacts expand the oracle's spatial and narrative scope. Ronny (tires), Gudrun (radio), and Rolf (drawer desk) anchor the interaction in different areas of the garage, encouraging visitors to move, touch, listen, and write. Together, these objects form a distributed system in which interaction unfolds across the space rather than at a single point.

We decided to give each object a voice that matched its attributed gender, character and also the content of the scripted prophecy. To this end, we divided the speaking roles among the team and also asked friends and neighbors for help, e.g., to be able to convey the specific local color of the Chemnitz Saxon dialect well. This way we not only created a collection of short WAV recordings of the artifacts stories but also of all scripted interaction of Ursula with the users and the additional hints to points of interest (POI) for the visitors. We recorded all text elements in German and translated them here.

We presented the artifacts along with their stories in some longer format than the audio records to tell a bit more about their background. We created bilingual printouts in German and English in typewriter style to present them in an authentic way. We placed the printouts a bit away from the artifacts on the right wall next near to the entrance to give interested visitors the opportunity to read them all. This way they do not distract from the artifacts themselves but visitors got the chance to read all the background stories.

We chose and integrated deliberately POI in Chemnitz that are linked local communities or a theme of garage and travel but are not so well known despite they deserve more attention – or in other words: some hidden gems: *Bagalut Garage* (another School of Garage), *Stadtfabrikanten Fablab* (a local maker space in an old industry building), *Chaos Computer Club* (home of the local club chapter with a small workshop), *Zietenaugust* (a community driven open gardening project), *Repair Café* (volunteers repair together with visitors broken devices), *Viaduct* (a steel arch railway bridge from the early 1900s showing craftmanship and elegance, *Stern-Garage* (one of first multi-story parking garages, built in the 1920s).

*female voice 1*

Welcome, I am Ursula, the Oracle of Chemnitz. To receive your prophecy,

Please tell me in which month you were born.

Your oracle prophecy is linked to… (I see your patron is …)

My name is **Ursula** – a telephone from the GDR of the 1970s. My aunt brought me home from the post office, where she worked, as if I were a treasure. But I was expensive: for almost fifty years she paid small installments for my plastic body and my heavy receiver. Every coin was like a promise that one day the voice of the world would reach her. Sometimes she imagined I was an oracle: dial the right number – and you don't just hear a voice, but perhaps the future itself. Today the smartphone laughs beside me – they carry entire fortune-telling apps within them. Yet whoever calls me still hears a faint crackle, as if the answer came from another time. I am Ursula – not a telephone of today, but an oracle of plastic, paid for with patience and longing.

In the GDR, there were only a few private telephone lines and, consequently, few telephones. It was something for people with privileges, connections, or important positions. Many households did not get a telephone connection until early 90s, and often then the first telephone was an older model that had been purchased cheaply, sometimes even a rotary dial phone. In the analog telephone age, the technology and appearance were very similar in Eastern and Western Germany. Despite the specific "Fe TAp 791-1" telephone is not an original GDR one it is very similar to the "Alpha Ferro" – in red too.

*male voice 2*

…is the Ronny tire, so please have a seat on it and stay on the phone. Mercedes seat will read your bottoms and help me to see your prophecy.

Hello! I am Ronny the Mercedes seat, born from the Bagalut community. I can see a young spirit in you – part Trabant soul, with an extra spark of creativity. Your energy spins like a wheel on asphalt and resonates like music from the garage.

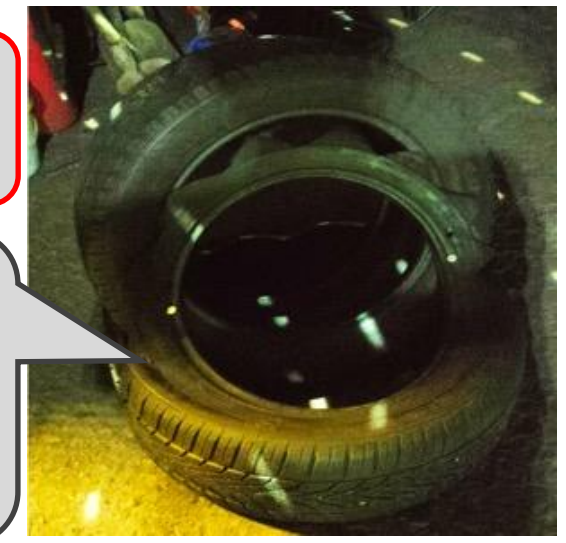

My name is **Ronny** – a seat made from old tires, born from the Bagalut community. My rubber once smelled of the road, but now carry the traces of asphalt and oil like memories of journeys from the past. I find energy in the people who lean on me and relax. Whoever sits on me can feel the rhythm of past travels and the heartbeat of new ideas.

Old tires are often found in garages and basements. In the economy of scarcity of the GDR, however, tires were valuable – even used ones. They were often "retreaded" because resources were scarce and to save money. Used tired were even imported from western countries to renew them in specialized companies for the domestic market. Nowadays, old tires are usually classified as hazardous waste. Nevertheless, old tires are often carelessly thrown away or unnecessarily hoarded in order to save on disposal costs. Thus, new and creative uses are an interesting option to give tires a second life.

*female voice 2*

…Erika, the typewriter. Type the first letter of your name on the typewriter…

Hello, I am Erika, a typewriter from the GDR. I once carried countless letters of love, longing, and resistance. In you I see the same determination – the ability to press forward, letter by letter, and shape your own future. Like the Chaos community, you are curious and playful with machines, seeking freedom of information and joy in discovery.

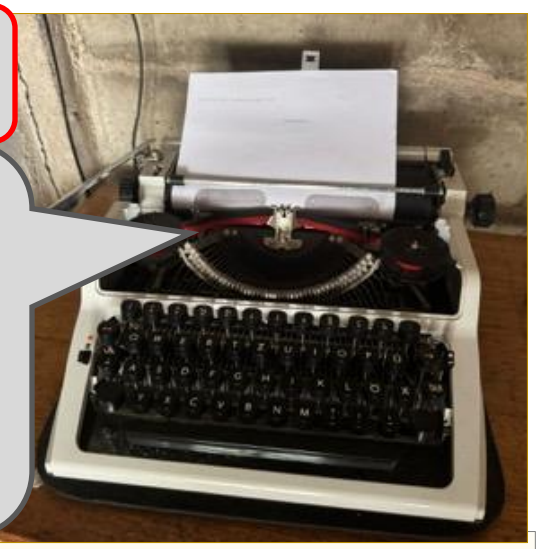

My name is **Erika** – a chic typewriter from the GDR. My body is light and stylish in light gray plastic. I happily clatered on the kitchen table, letters jumped onto the paper like little soldiers. Mistakes? No "back" button, only Tipp-Ex and patience. Then came the computers – they were slimmer than me and write silently. They are also more colorful and can erase mistakes for free with every typo. But now we live next to each other in the Computer Club Chemnitz and we get along quite well. When someone touches me, they feel the heartbeat – I am not a machine of today, but a reminder that every word once had its price.

The former East German typewriter brand "Erika" was produced in Saxony and Thuringia, some models even in Chemnitz. It was the most widespread typewriter in Germany with over eight million units sold over decades, starting with early mechanical ones up to electronic ones till the early 1990s. Most people over the age of 40 in Eastern Germany know the brand. The typewriters were also sold cheaply for export to western countries, bringing much-needed foreign currency into the GDR. Erika typewriters are known for their high-quality print and durability - some people still love and use them.

*female voice 3*

…the radio Gudrun. To hear your prophecy choose any frequency that feels right to you. Then stay on the line and wait for Gudrun's voice.

Hello, I am Gudrun, a radio from another time. From the frequency you tuned, I can already read your vibration… – you are both a source of knowledge and of joy. You can tune people in, bridge distances with your voice, and turn silence into connection.

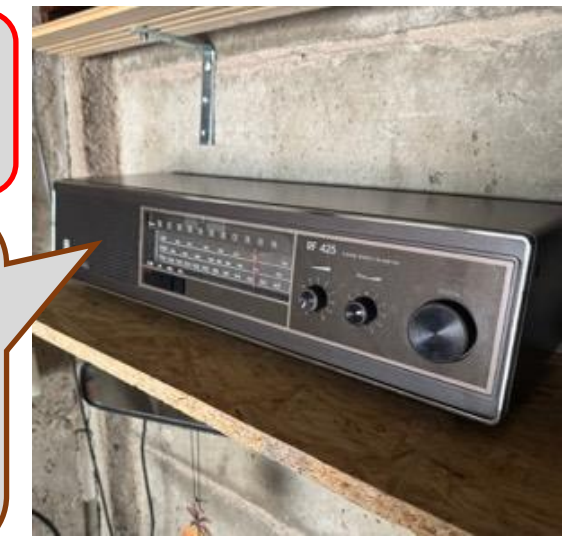

My name is **Gudrun** – a radio born in a time when every crackle in the air carried a promise. Turn my dial, and not only does a station begin to speak, but also an invisible bridge grows – between rooms, cities, sometimes even worlds. From my speakers once came forbidden waves, but also music, news, and weather reports. Then everything spun like a whirlwind: first came cassettes, then CDs, later MP3s... and at last I was left in the garage. Yet whoever listens to me today can still hear the echo of another time.

The RFT combine dominated the radio and television market in the GDR. Although color televisions were less common in the East, analog radios were also widespread there and similar between East and West. These old devices still have their fans, who are deliberately interested in the aesthetics and also the old, imperfect but as warm and full considered Lo-Fi sound – as a contrast to supposedly ever-improving (digital but sterile) Hi-Fi. There is an active community of collectors specifically for RFT devices who exchange devices and restore them at great expense to keep them in working order.

*male voice 1*

…Manfred, the mixer. Put your hand inside the mixer for two seconds… but stay on the line and wait for his voice.

Hello, I am Manfred, the mixer from an old kitchen. I was made to spin, to stir, to bring together different tastes and textures into something new. In you I sense the same talent – you know how to blend people, ideas, and moments into harmony. You are not afraid of a little noise or mess because you know creation needs motion.

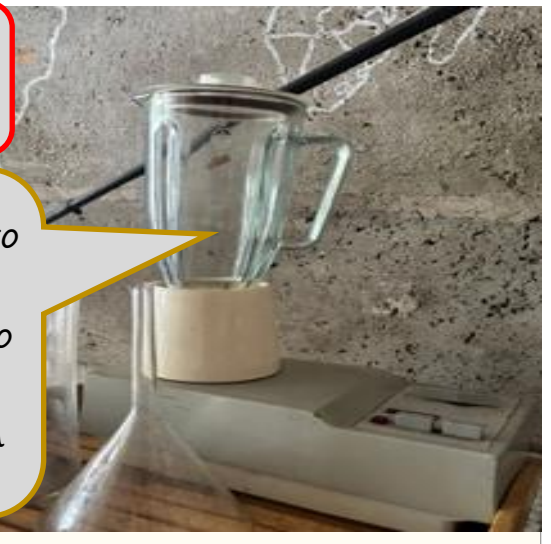

My name is **Manfred**. Built in 1974, gray, with a powerful motor. A heavy, reliable blender. My class container is thick, my blade sharp, built to chop anything that comes my way – onions, nuts, ice cubes... creating new tastes and textures. I danced for hours in the kitchen – my noise mingling with laughter and radio music. Over the years, new appliances came along, shinier, quieter, smaller. I was moved to the back of the cupboard, then to the garage. A little dusty, but not forgotten. Today when I'm brought out, my motor may whir more heavily, but my heart still beats to the beat of the seventies.

The multi-purpose kitchen machine "KM8" or the famous "AKA electric RG28" were very popular appliances in the GDR. Due to the planned economy, the range of consumer goods was limited and some devices remained unchanged on the market for decades. Accordingly, there is a good chance that one may have used such a device or at least seen it in use by parents or grandparents. The devices have a reputation for being very robust and nearly indestructible. Accordingly, some models are even still in use despite being decades old, are still in demand and are traded on the second-hand market.

*male voice 3*

…Rolf, the drawer from a 1970s desk. Inside there is a book filled with visitors' words. Open the top drawer and read aloud the first short entry that catches your eye… then hurry back to the phone, because Rolf can't wait to tease out your energy.

Ha! So here you are. I am Rolf, the drawer – stuffed with secrets, doodles, and cookie crumbs from the 70s. The words you just read? They bounce around inside you like a ping-pong ball. Strong and steady like the old railway bridge, but also quick and light – ready to play, to laugh, to connect people on both sides of life's table.

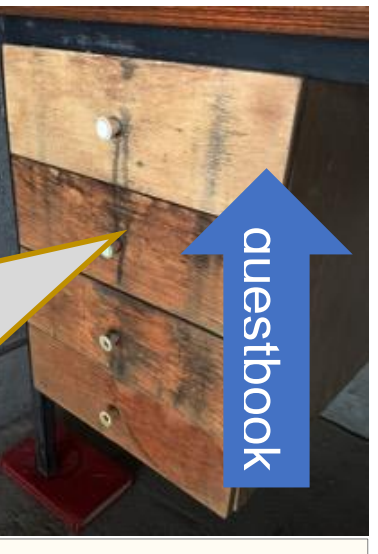

They called me **Rolf** – a drawer in an old table. Inside me lived little notes, secret letters, and forgotten crumbs. Once I stood in a children's room with a view of the park, wishing I could go out and play in the sun. Instead, I ended up in the dark garage. But open me today, and you'll still find the children's laughter and the rustle of tiny secrets.

Similar to kitchen appliances, furniture in the GDR was also produced under a planned economy in combines (large central led industrial conglomerates). These furniture combines specialized in specific lines and models, which they produced for the entire country. As a result, some models, such as children's furniture and desks or whole model series, e.g. the "602", were widely known throughout the country – similar to today's IKEA classics such as the Billy bookshelf. People still like the style of these old GDR furniture, often in practical and clean design, some even consider iconic.

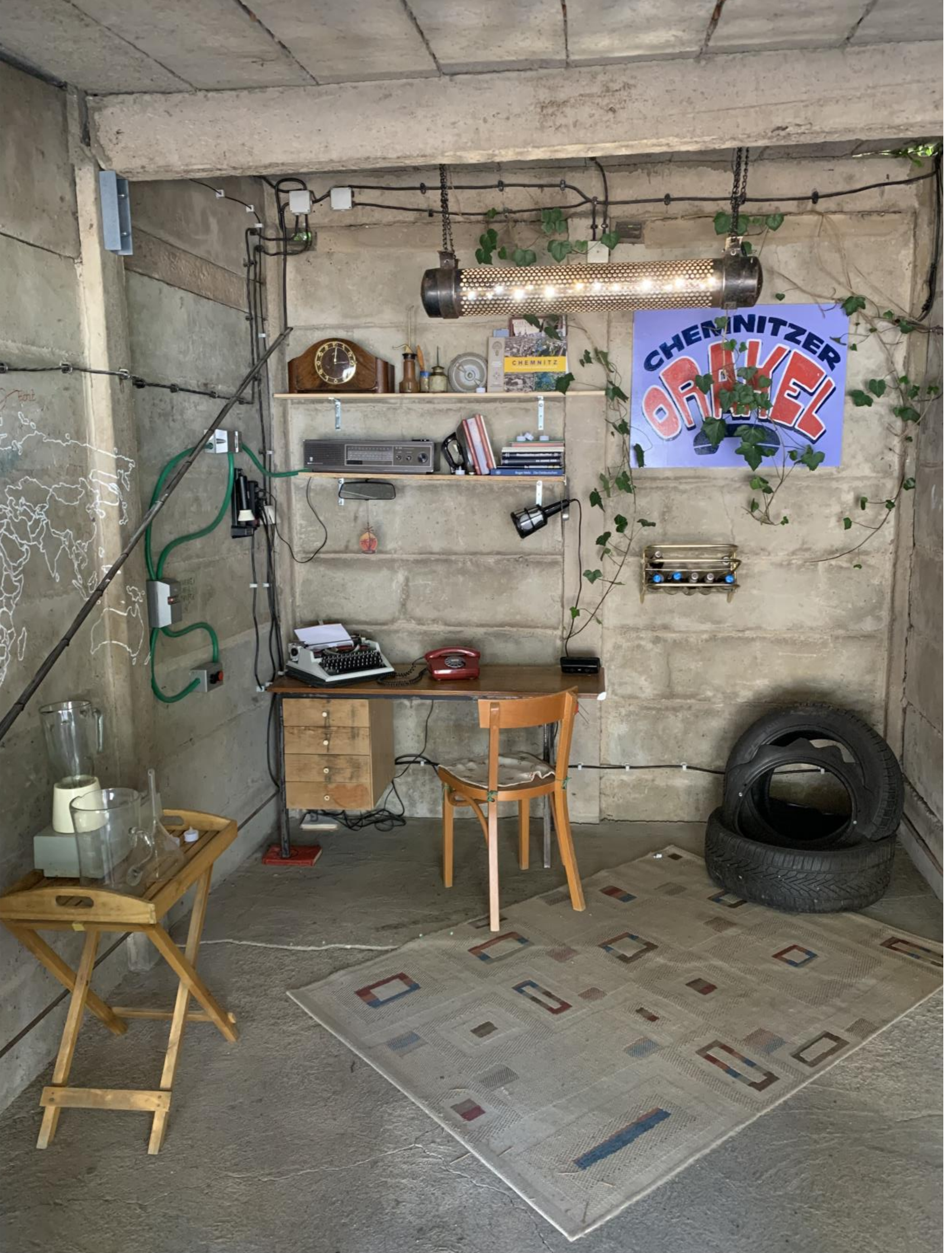

### The installation ready for the public opening

For the final pop-up installation we cleaned out the stuff again. The installation consists not only of the selected artifacts but also other decorative stuff and the context of the garage itself. We aimed to create a natural garage feeling what took a lot of effort.

The artifacts were placed at the back of the garage. Rolf the drawer-desk found its place in the left corner together with Erika the typewriter and Ursula the telephone oracle on top of it. A small serving side table along the left wall is for Manfred the mixer. Ronny the old tires found their place right next to the desk in the right corner. We arranged and installed the other salvaged decorative materials such as a shelf for Gudrun that also homes an old clock, some oil cruets, an old flat iron and some books about Chemnitz, East Germany and the East Germans. A heater cartridge is hanging from the ceiling and an old carpet is placed together with a chair in front of the desk. The oracle sign was place right next to the shelf to greet entering visitors. The PIR sensor was placed hidden behind the concrete beam on the ceiling and adjusted to register movement when crossing the carpet on the floor half-way. This installation allowed that the sensor was triggered by any approach towards the desk or chair in the corner. Even the beer rack right next to the desk is a deliberately important part of the installation. Having a good time, enjoying the garage for fun, hobby and relaxation is an important part of the garage culture – what often involves also a beer – alone or together with fellow garage owners.

On the left side of the garage three painted maps show the contours of Europe, Germany and the world, and invite visitors to take a paint marker and to leave a marking where they are from. This way the visitors can bring in slowly but cumulatively some color to the otherwise still grey and bleak walls.

Although the garage had now been tidied up and visually transformed into a pop-up installation, there were still many remnants of its past and the garage still has this special atmosphere. This includes the typical unpainted, rough, window-less walls made of preformed concrete parts, the cramped confines of the narrow width and low concrete ceiling, the imperfect concrete floor, and the light coming in only through the open garage door in the back of anybody entering. The unmistakable smell, similar to that of old cellars, hung still in the air, characterized by a hint of stagnant moisture, oil, and gasoline. This smell is also embedded into the salvaged items that were stored in garages and which developed some patina of slight mold stains and rust and that carry on and emit themselves this smell for a long time. Some visitors may even be reminded of the two-stroke exhaust fumes of a Trabant, the iconic GDR type car that used to be parked here or the characteristic whining sound of such an engine somewhere in the distance on the garage yard.

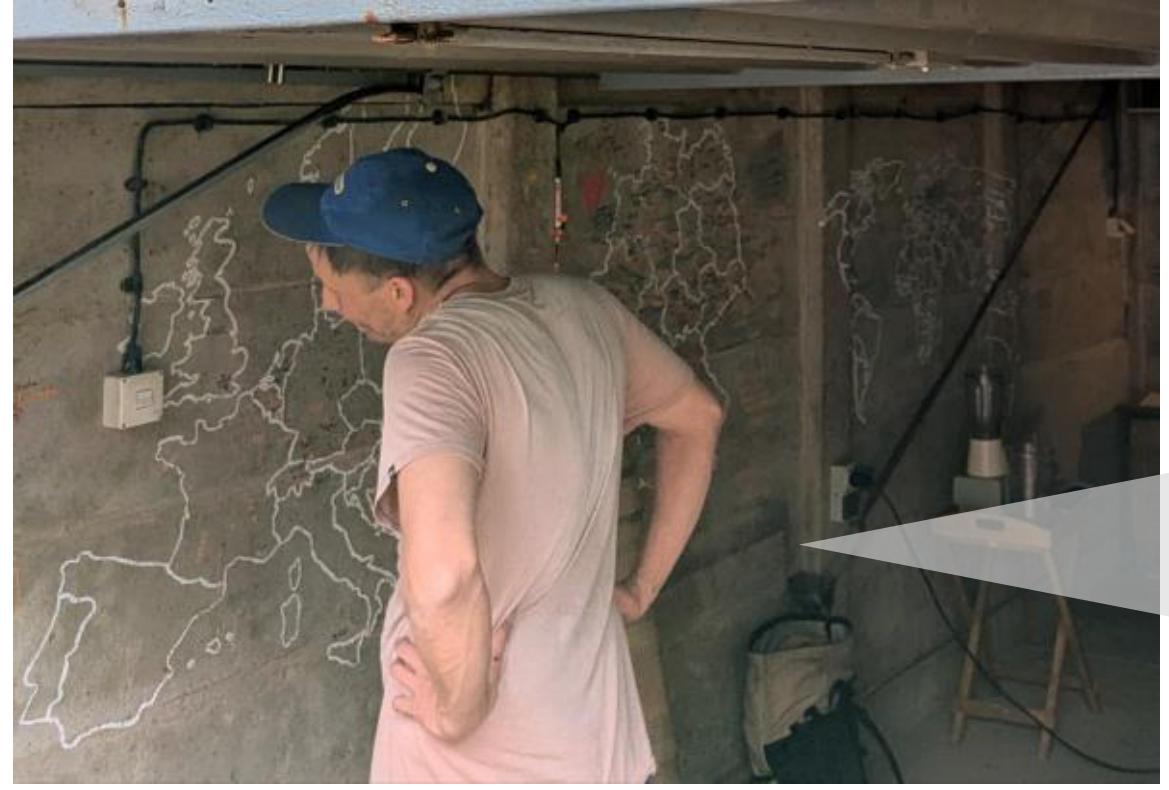

The visitors appreciated the maps on the wall and created a permanent trace of their visit. The dots show visitors from all across Germany, Europe and even the world.

## Visitors' reception and interactions

Visitors arrived at the garage typically alone or in pairs, and sometimes in small or even large groups. We registered after opening in August 2025 around 2.700 visitors interacting with the system before closing for Winter break in the end of November.

During the opening weekend members of the garage team were present for interaction with the visitors. In the following weeks the garage was kept open by the garage owner and neighbors who volunteered to keep an eye on the installation. Local members of the garage team visited the installation from time to time. When present, we observed how visitors interacted with the installation and took brief notes and (after agreement) some photos.

The conceptualized interactions were usable and accessible so that visitors of all background and age were able to interact with it. The setup worked well without any major problem and kept operational during the complete time. Having more a 3rd wave HCI lens (Bødker, 2006; 2015) on everyday lives and culture we focused following on less usability or the human factor but rather on how visitors experienced the installation in its context and what effect this had on them.

Overall, the visitors were keen to interact with the installation and enjoyed it very much. Some visitors reacted promptly to the ringing telephone, others were initially irritated and shy and even moved back what caused that the ringing stopped. Eventually the majority of visitors was curious and furious enough to pick up the phone and to interact with the installation. Often one visitor interacted with the installation, observed by another visitor, and reported then names and parts of the story to other visitors. Shy visitors observed this first and then became curious on their own and eventually also interacted with the installation.

Some of the visitors tried the oracle only for one story, other visitors for more. Some visitors were even keen to grasp all stories what required several attempts and quite some persistence given the random selection of a story. Ursula the telephone was the star and center of the installation but also Gudrun and Erika were favorites. Manfred was also favored - despite a bit more fragile. Even Ronny got some attention and was used in the indented way as a comfortable seat. Rolf as the keeper of the guestbook was also regularly involved.

Sehr schön!! Gut gemacht Ursula! Auch wenn
du mir nicht geantwortet hast.

The **guestbook** provided us with feedback and messages commenting about Chemnitz, garages, the #3000 garages and School of Garage project as well as the Oracle of Chemnitz. We selected and translated some of the mostly in German created entries to illustrate the feedback we got. Overall, the visitors reacted very positive. Some entries in the guest book are even enthusiastic or illustrated even with smileys or hearts.

*We, two friends, visited their first garage and are thrilled for all the details. The phone scared us at first, then delighted us. 1000 thanks and stay creative, CHEMNITZ!!!*

*Thanks for the oracle and for exploring the potential of garage culture. There's definitely more to come!*

Visitors liked the framing, and in some cases, it brought back **their own experiences and memories**.

*Wonderful—just like in the old days*

*Great garage. We had fun. All I need to do now is get myself a Trabi! Thanks to the oracle for the message of happiness. [remark: "Trabi" is colloquial for "Trabant", the most iconic East German car from 1950s-80s]*

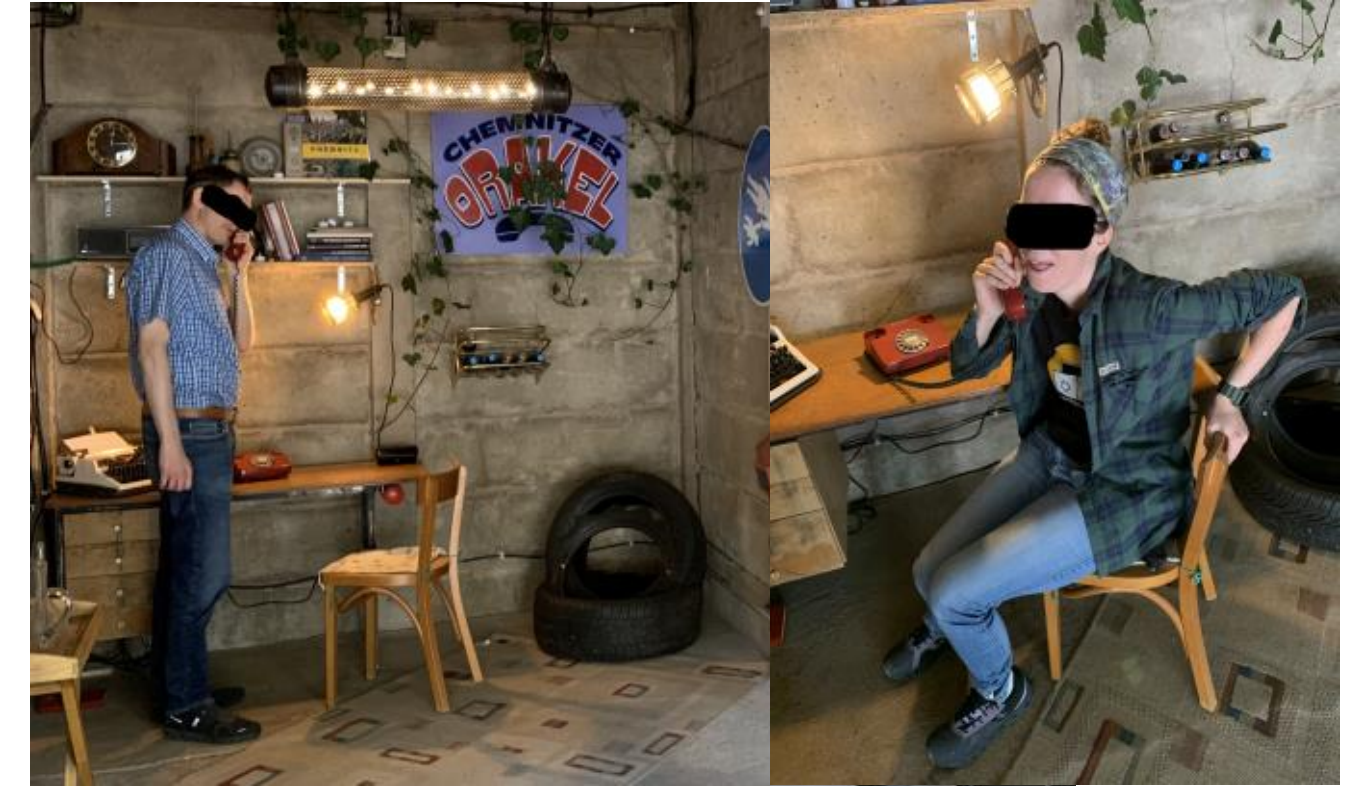

Some visitors came even for a **second or even third time** and also documented this in the guestbook.

*We're back again and find your garage soooo original and charming.*

Some of the visitors directed their guestbook entry to the oracle, indirectly to the **involved artifacts** or even directly to them by **naming them**.

*This is a garage showing a really well done idea. Creative! Greetings to Ursula, Rolf + Mannie [Mannie is a colloquial form of Manfred in German]*

*Very nice. Well done, Ursula! (Even though you didn't answer me.)*

*A nice exhibition + thanks Gudrun!*

*Thank you, Ursula, you provided me very much joy.*

The visitors appreciated not only the interactive parts of the system but also the small **details of the decoration**.

*What a beautiful garage with so lovely books, especially the illustrated book 1926-2016. You have lovingly created something here that brings us closer to life in Chemnitz. Thank you very much.*

The visitors also the mentioned the provided bottles of **beer** as some lovely detail, very welcomed especially in the first still hot late summer weeks.

*A very cool project and thanks for the beer!*

*Thank you very much for the trustfully provided stash of beverages!*

In addition to the guestbook we also screened **social media postings** which were in a similar positive sentiment as the guestbook entries.

## The creators' perspective

Some local voice of a neighbor, drawn into the ongoing project and becoming active part of the garage team, expresses what the Oracle of Chemnitz meant for the participants and how the project initiated a local impact:

*"…suddenly you're part of their project and spend the day with lovely, smart and creative folks you probably never have met otherwise. It was so much fun being able to support this project with my passion for words and even lending some parts my voice. And now we have our very own little oracle right by our house - inviting visitors und curious folks to explore our home town in a very silly und fun way."*

All the fun and positive energy in the process culminated in a great interactive system conceptualized and created out of nothing within one week. All the positive reactions and voices we got from the visitors fill us with warmth. Accordingly, the team decided to rename from HackLab garage to **HackLuck garage**.

## Discussion

The Oracle of Chemnitz illustrates how creativity and change can emerge from everyday places through collaborative making and the reactivation of familiar technologies. Rather than framing innovation as a laboratory-driven or purely technological process, the project situates creative practice within a historically and socially charged environment – the garage.

The vignettes collected through the guest book show that visitors engaged not only with the interactive system itself, but also with atmosphere, materiality, and social cues. Familiar objects such as a rotary-dial telephone, a typewriter, or a beer rack supported encounters that were playful, reflective, and socially situated. These interactions suggest that creativity was experienced as something embedded in shared time, informal exchange, and recognizable practices, rather than as a distant or abstract concept.

Retro-tech and tech-nostalgia such as in the telephone or the typewriter arouse curiosity, both through familiarity and through unexpectedness, especially when having a relation to the specific context such as their origin in the Eastern Bloc.

The artifacts and their stories worked as conversation pieces (Erlhoff & Marshall, 2008) that successfully triggered reports of own experience and memories, similar to encounters reported in (Köpferl & Kurze, 2025a, 2025b).

The installation manages to connect different types of human-technology relations. Visitors interact with the technology on site (alterity relation), which reveals previous uses and significance in their everyday life of former days, associated stories about it etc. (background relations) (Ihde, 1990). The oracle successfully picks up participatory approaches that are speculative as well as embedded in their context (Berger et al., 2017; Mucha et al., 2022). The format of the summer school proved to be very efficient in terms of the achieved result and participation. As participation format this go beyond what even a Living Lab approach can achieve (Bischof et al., 2018, 2020). The highly participative character of the format shows exceptionally well in the involvement of neighbors far beyond initial planning.

As a situated case, the Oracle of Chemnitz demonstrates how the abstract goals articulated in the Chemnitz2025 cultural strategy—such as fostering a culture of making and self-efficacy—can take material form in small-scale, everyday settings (City of Chemnitz, 2018).

The project can also be seen as an alternative to the much-praised garage as a symbol of Silicon Valley startups (Meisenzahl, 2020) and their mentality of creativity, founding ideas and determination (Godelier, 2007), of new beginnings and innovation but sometimes also disruptive change (Taplin, 2018). Our garage demonstrated that a garage spirit of **moving fast and repairing things** also works and results in successful, enjoyable techno-artistic outcomes. The concept of the garage as a creative laboratory for ideas, exchange, and collective action - without a business plan but a shared vision - as well as the reuse and mashup of old things for creating something enjoyable with reducing waste instead of generating new one underline this.

## Conclusion

Looking back at the #3000Garagen initiative in general, the School of Garage in particular, and the HackLab garage specifically, we consider the project to have been highly successful in terms creativity and change in the process, collaboration, and public engagement.

The creative process – often marked by improvisation and some moments of productive chaos – combined interdisciplinary and transdisciplinary collaboration with active knowledge exchange and the involvement of local communities and neighborhoods. These elements came together in the conception and realization of a fully functioning interactive installation, developed collaboratively within a single week.

The Oracle of Chemnitz integrated the history, spirit, and sensory qualities of garage culture into a coherent artistic and interactive installation that attracted a large and diverse set of visitors. The installation enabled visitors to (re)discover aspects of Chemnitz's history, East German garage culture, and, in many cases, their own personal memories. More broadly, the project suggests that creative cooperation and collaborative making can inspire renewed forms of garage culture – grounded in shared practices, openness, and collective experimentation.

As a pictorial contribution, this work does not aim to generalize its findings. Instead, it presents a situated case study demonstrating how interactive installations can draw on local cultural practices to foster engagement and meaning. In doing so, the Oracle of Chemnitz contributes to ongoing discussions by showing how creative experiences may be shaped through place, material history, and collaborative processes in everyday settings.

## Acknowledgment

The project was independently organized and not university funded. We heartily thank the team of Constructlab, the summer school participants, the garage provider as well as Chaos Computer Club Chemnitz for their initiative and support. The text was proofread using the online AI-service DeepL.

Resize>Remix>Regen Th/ngs!

## References

Adams, J., Dedehayir, O., O'connor, P., & Nokelainen, T. (2019). Who are the adopters of retro-technology? *International Journal of Innovation Management*, *23*(08), 1940009. https://doi.org/10.1142/S1363919619400097

Berger, A., Totzauer, S., Lefeuvre, K., Storz, M., Kurze, A., & Bischof, A. (2017). Wicked, Open, Collaborative: Why Research through Design Matters for HCI Research. *I-Com*, *16*(2), 131–142. https://doi.org/doi:10.1515/icom-2017-0014

Bischof, A., Freiermuth, M. M., Storz, M., Kurze, A., & Berger, A. (2020). » Living Labs «als Beispiel für die konzeptionellen Herausforderungen der Integration von Menschen in Technikentwicklung. In *Das geteilte Ganze: Horizonte Integrierter Forschung für künftige Mensch-Technik-Verhältnisse*. Springer. https://doi.org/10.1007/978-3-658-26342-3

Bischof, A., Kurze, A., Totzauer, S., Storz, M., Lefeuvre, K., & Berger, A. (2018). Initiating Participation: Methodological and Practical Challenges of Living Lab Projects for Early Stages of Research and Development. *Conference Proceedings Open Living Lab Days 2018*, 407–421. https://doi.org/10.5281/zenodo.1434972

Bødker, S. (2006). When second wave HCI meets third wave challenges. *Proceedings of the 4th Nordic Conference on Human-Computer Interaction: Changing Roles, NordiCHI '06*, 1–8. https://doi.org/10.1145/1182475.1182476

Bødker, S. (2015). Third-wave HCI, 10 years later—Participation and sharing. *Interactions*, *22*(5), 24–31. https://doi.org/10.1145/2804405

Bong, W. K., Maußer, F., van Eck, M., De Araujo, D., Tibosch, J., Glaum, T., & Chen, W. (2020). Designing Nostalgic Tangible User Interface Application for Elderly People. In K. Miesenberger, R. Manduchi, M. Covarrubias Rodriguez, & P. Peňáz (Eds.), *Computers Helping People with Special Needs* (pp. 471–479). Springer International Publishing. https://doi.org/10.1007/978-3-030-58805-2_56

Bonvin, L. A. (2023). *Echoes of the East: Unearthing (Post)Socialist Heritage in Chemnitz' Garages*. https://urbanstudies.philhist.unibas.ch/fileadmin/user_upload/urbanstudies/Dokumente/StudentWork/2403_Bonvin_Master_Thesis.pdf

Campopiano, J. (2014). Memory, Temporality, & Manifestations of Our Tech-nostalgia. *Preservation Digital Technology & Culture*. https://doi.org/10.1515/pdtc-2014-0004

City of Chemnitz. (2018). *Bid Book 2: "C the unseen"* (1st ed.). https://chemnitz2025.de/fileadmin/khs/03_INFORMIEREN/Bidbook/BidBook2_Chemnitz2025_english.pdf

Constructlab. (2025). *The School of Garage Chemnitz 2025—Constructlab—Transnational Collaborative Network*. https://constructlab.net/school-of-garage-chemnitz-2025/

Crowley, D., & Reid, S. E. (Eds.). (2002). *Socialist Spaces: Sites of Everyday Life in the Eastern Bloc*. Berg Publishers.

Dunne, A., & Raby, F. (2013). *Speculative Everything: Design, Fiction, and Social Dreaming*. MIT Press.

Erlanger, O., & Govela, L. O. (2018). *Garage*. MIT Press.

Erlhoff, M., & Marshall, T. (Eds.). (2008). CONVERSATION PIECE. In *Perspectives on Design Terminology* (pp. 79–79). Birkhäuser. https://doi.org/doi:10.1007/978-3-7643-8140-0.79b

Frankjær, R., & Dalsgaard, P. (2018). Understanding Craft-Based Inquiry in HCI. *Proceedings of the 2018 Designing Interactive Systems Conference, DIS '18*, 473–484. https://doi.org/10.1145/3196709.3196750

Godelier, E. (2007). "Do You Have a Garage?" Discussion of Some Myths about Entrepreneurship. *Business History Conference. Business and Economic History On-Line: Papers Presented at the BHC Annual Meeting*, *5*, 1.

Ihde, D. (1990). *Technology and the lifeworld: From garden to earth*. Indiana University Press.

Kamiński, R., & Rosłoń, R. (2024). A small garage on the sidelines of a corporation: Should innovative projects be isolated? *Ekonomia*, *29*(3), 7–17. https://doi.org/10.19195/2658-1310.29.3.1

Kang, L. (Leo), Jackson, S. J., & Sengers, P. (2018). Intermodulation: Improvisation and Collaborative Art Practice for HCI. *Proceedings of the 2018 CHI Conference on Human Factors in Computing Systems, CHI '18*, 1–13. https://doi.org/10.1145/3173574.3173734

Kastner, J. (2019). The Domestication of the Garage. *Places Journal*, (2019). https://doi.org/10.22269/190205

Köpferl, K., & Kurze, A. (2025a). Human-Chatbot Interaction: When ChatGPT meets an old Typewriter. *Proceedings of the 21st International Conference on Culture and Computer Science: From Humanism to Digital Humanities, KUI '24*, 1–8. https://doi.org/10.1145/3719236.3719250

Köpferl, K., & Kurze, A. (2025b). Write Again(st) the Machine. Reanimating a GDR-Era Typewriter as a Reflective Interface for Human-AI Dialogue. *Mensch Und Computer 2025 - Workshopband*. https://doi.org/10.18420/muc2025-mci-demo-280

Krause, G. (2025). *Kulturhauptstadt: Chemnitz macht die Garage zum Kulturschatz | MDR.DE*. https://www.mdr.de/nachrichten/sachsen/garagen-kulturhauptstadt-ddr-kultur-news-100.html

Kulturhauptstadt Europas Chemnitz 2025 gGmbH. (2025). *Chemnitz 2025: #3000Garagen*. https://chemnitz2025.de/en/3000garagen/

Meisenzahl, M. (2020, April 1). *Starting in a garage is crucial to the origin story of many Silicon Valley entrepreneurs. Here are the modest beginnings of 5 tech companies worth billions today.* Business Insider. https://www.businessinsider.com/google-apple-hp-microsoft-amazon-started-in-garages-photos-2019-12

Mucha, H., Barros, A. C. de, Benjamin, J. J., Benzmüller, C., Bischof, A., Buchmüller, S., Carvalho, A. de, Dhungel, A.-K., Draude, C., Fleck, M.-J., Jarke, J., Klein, S., Kortekaas, C., Kurze, A., Linke, D., Maas, F., Marsden, N., Melo, R., Michel, S., … Berger, A. (2022). Collaborative Speculations on Future Themes for Participatory Design in Germany. *I-Com*, *21*(2), 283–298. https://doi.org/10.1515/icom-2021-0030

Seo, J. A., Cho, H., Lee, S., & Cheon, E. (2025). Back to the 1990s, BeeperRedux!: Revisiting Retro Technology to Reflect Communication Quality and Experience in the Digital Age. *Proceedings of the 2025 CHI Conference on Human Factors in Computing Systems, CHI '25*, 1–19. https://doi.org/10.1145/3706598.3713568

Taplin, J. (2018). *Move Fast and Break Things: How Facebook, Google and Amazon Have Cornered Culture and Undermined Democracy*. Pan.

Tuvikene, T. (2011). Nõukogude garaažikultuur. Soviet Garage Culture. *Methis. Studia Humaniora Estonica*, *5*(7). https://doi.org/10.7592/methis.v5i7.539

Wehner, M. (2025). *Chemnitz ist Kulturhauptstadt Europas 2025: Seine heißbegehrten Garagen*. FAZ.NET. https://www.faz.net/aktuell/stil/drinnen-draussen/chemnitz-ist-kulturhauptstadt-europas-2025-seine-heissbegehrten-garagen-110362641.html